\documentstyle[11pt,eclepsf,aaspp4]{article}
\begin{document}
\newcommand{\simgt}{\lower.5ex\hbox{$\; \buildrel > \over \sim \;$}}
\newcommand{\simlt}{\lower.5ex\hbox{$\; \buildrel < \over \sim \;$}}
\title{Measuring Cosmological Parameters with the SDSS QSO 
Spatial Power Spectrum Analysis to Test the Cosmological Principle}
\author{Kazuhiro Yamamoto\altaffilmark{1}}
\affil{ 
        Max-Planck-Institut for Astrophysics,
                Karl-Schwarzschild-Str. 1,
                D-85741 Garching, Germany
\\
Department of Physical Science, Hiroshima University,
        Higashi-hiroshima, 739-8526, Japan
}
\date{ in original form 2002 May 20
}
\altaffiltext{1}{e-mail: kazuhiro@hiroshima-u.ac.jp}

\begin{abstract}

In this paper we emphasize the importance of the Sloan 
Digital Sky Survey (SDSS) QSO
clustering statistics as a unique probe of the Universe.
Because the complete SDSS QSO sample covers a quarter of the
observable universe, cosmological parameters estimated
from the clustering statistics have an implication as a 
test of the cosmological principle, by comparing with 
those from local galaxies and other cosmological observations.
Using an analytic approach to the power spectrum for the 
QSO sample, we assess the accuracy with which the 
cosmological parameters can be determined.
Arguments based on the Fisher matrix approach 
demonstrate that the SDSS QSO sample might have 
a potential to provide useful constraints on 
the density parameters as well as the cosmic 
equation of state.\\
\vspace{1mm}
\end{abstract}

\vspace{0.3cm}

\section{Introduction}

The cosmological principle, the assumption of the homogeneity
and isotropy of the Universe on large scales, is one of 
the most fundamental in the framework of the cosmology
(e.g., Peebles 1993).
As Lahav (2001,2002) reviewed, observations of the comic 
microwave background (CMB), the cosmic X-ray background, 
radio sources, and the Lyman-$\alpha$ forest strongly support 
the cosmological principle, though it is difficult 
to prove the principle definitely. 
Actually inhomogeneous cosmological models, which challenge 
the cosmological principle, have been 
proposed (Kantowski \& Thomas 1998, Kantowski 2001, 
Barrett \& Clarkson 2000, Celeria 2000).
For example, it is claimed that a large local void universe 
might be viable, which is compatible with the observations 
of type Ia supernovae, the cosmic microwave isotropy, 
and the distribution of the local galaxies (Tomita 2000,~2001).

We emphasize the importance of the QSO clustering statistics 
as a unique and independent probe of the Universe. 
Especially it provides a chance to test the cosmological 
principle by comparing the cosmological parameters measured
from the QSO clustering analysis with those from local 
galaxies and other cosmological observations. 
Recently the two degree field (2dF) QSO redshift (2QZ) Survey 
reported that a simply biased mass distribution explains 
the QSO spatial clustering and that the cosmological 
parameters can be measured (Croom et~al. 2001, 
Hoyle et~al. 2002a; 2002b, Outram et~al. 2001).
Their preliminary results with the $10$k catalogue favor 
the cold dark matter (CDM) model with a cosmological constant, 
though the constraint is not very tight (see also Yamamoto 2002; 
Hereafter Paper I).

On the other hand, the Sloan Digital Sky Survey (SDSS) will aim 
to compile a homogeneous catalogue of $10^5$ QSOs as well as 
$10^6$ galaxies. The SDSS stores the 5 bands CCD imaging of $10^4$ 
deg${}^2$ in the North Galactic Cap and of three $2.5\times 90$ 
deg${}^2$ strips in the South Galactic Cap. The QSO spectroscopic
survey of the North Galactic Cap will detect $10^5$ QSOs in the
range of the redshift $0\simlt z \simlt 5$ with the limiting
magnitude $m=19\sim 20$. A QSO catalogue with the SDSS early 
data release has been reported (Schneider et al. 2002).
Because the SDSS QSO sample will cover a quarter of the
observable Universe, the cosmological parameters 
measured with the clustering statistics provide a 
unique implication for the cosmological principle. 
Therefore it is important to estimate the accuracy 
with which the QSO clustering statistics can measure 
the cosmological parameters.

Several authors have proposed possible methods (e.g., 
Ballinger, Peacock, \& Heavens 1996, Matsubara \& Suto 1996, 
Popowski et~al. 1998),
which originate from the Alcock-Paczynski's test (1979).
Very recently, Calvao et~al. (2002) have claimed that a significant 
constraint on the density parameters as well as the cosmic 
equation of state of the dark energy, which is a quite
important issue in physics and cosmology~(see e.g., Caldwell,
Dave \& Steinhardt 1998), can be obtained from 
the Alcock-Paczynski test with the 2dF QSO sample. 
Concerning the SDSS QSO sample, however, a systematic assessment 
on the capability of measuring the cosmological parameters 
have not been well investigated, as far as we know.
The reason might come from the fact that physical processes 
of the QSO formation and time-evolution have not been 
completely understood. On the other hand, the number count of the SDSS 
QSO sample in the early data release has been demonstrated
by Schneider et~al.~(2002), which we use in our investigation. 

Motivated by these situations, we revisit the QSO clustering 
statistics and assess the accuracy with which the cosmological 
parameters can be measured by the SDSS QSO sample.\footnote{
Almost this work was completed, a similar paper by Matsubara 
and Szalay (2002) has been announced. Some differences are 
mentioned in the last section.}
For this purpose we adopt the Fisher-matrix approach, which has been  
often used to assess the accuracy of measurement of the cosmological
parameters from large surveys such as CMB anisotropy 
observations, galaxy redshift survey, and supernova data sets.
The Fisher matrix approach bases the method on mathematical 
theorems, which enables us to clearly interpret results.
The method adopted in the present paper is a simple 
extension of an approximate Fisher matrix approach 
using the power spectrum, which was originally formulated 
for galaxy samples (Tegmark et~al. 1998; Tegmark 1997).
We use the power spectrum analysis extended so as to
incorporate the various observational effects, the light-cone 
effect, the linear redshift distortion and the geometric 
distortion, which is useful for QSO samples (Paper I,
Suto et~al. 2000, Yamamoto \& Suto 1999, Yamamoto \& Nishioka 2001).
This paper is organized as follows: In section 2, we
briefly review formulas to evaluate the Fisher matrix element. 
In section 3, the best statistical errors of the cosmological 
parameters are shown assuming the complete SDSS QSO sample, which
demonstrates the usefulness of the power spectrum analysis
of the sample. In section 4, the problem of degeneracy
in the parameter space is briefly discussed.
Section 5 is devoted to conclusions.
Throughout this paper we use the unit in which 
the light velocity equals 1. 

\section{Method}
\newcommand{\cpara}{c_{\scriptscriptstyle \|}}
\newcommand{\cperp}{c_{\scriptscriptstyle \bot}}
\newcommand{\qpara}{q_{\scriptscriptstyle \|}}
\newcommand{\qperp}{q_{\scriptscriptstyle \bot}}
\def\bfs{{\bf s}}
\def\bfk{{\bf k}}
\def\barn{{\bar n}}
\def\calF{{\cal F}}
\def\tr{{}}

The Fisher matrix approach provides a useful technique 
to estimate errors in measuring cosmological parameters
from a given data set (see e.g., Tegmark, Taylor \& Heavens 1997,
Jungman et~al. 1996;1997). 
Tegmark et~al. (1998) developed a useful approximate method 
to estimate the Fisher matrix element using the power 
spectrum of large surveys of galaxies. 
The method used in the present paper is a simple extension 
of their work. 
In an analysis of the QSO clustering, the additional 
observational effects, the light-cone effect and the 
geometric distortion effect, must be taken into account.
The power spectrum analysis incorporating these effects
can be developed by extending the work by 
Feldman, Kaiser \& Peacock (1994; hereafter FKP).
We start from briefly reviewing this power spectrum analysis
useful for the QSO samples.

First we assume that the number density field of sources 
$n(\bfs)$ is constructed from a data set, where $\bfs$ 
denotes the three dimensional position of the sources.
Here we assume that a distance-redshift relation 
$s(=|\bfs|)=s(z)$ was adopted to plot the map (see below for the 
explicit definition $s(z)$).
Following FKP, we define the density fluctuation field 
by 
\begin{eqnarray}
F(\bfs)={n(\bfs)-\alpha n_s(\bfs)
\over[\int d\bfs \barn(s)^2]^{1/2}},
\end{eqnarray}
where $\alpha$ is a constant, $n_s(\bfs)$ is a synthetic catalogue,  
$\barn(s)$ is the expected mean number density. 
The optimum weight factor introduced in FKP can 
be fixed as a constant because $\barn P(k)$ is 
less than unity for the QSO sample. This means that
the power spectrum analysis considered in the present 
paper does not require information on the evolution 
of the power spectrum as a function of the redshift. 
The power spectrum is constructed by the square of
the Fourier coefficient of the fluctuation field
defined by 
\begin{eqnarray}
\calF(\bfk)=\int d\bfs F(\bfs) 
e^{i\bfk\cdot \bfs}. 
\end{eqnarray}
Then the estimator of the power spectrum
(before averaging over a shell of the radius $k(=|\bfk|)$
in the Fourier space) is defined by subtracting the shot 
noise contribution $P_{shot}$ from $|\calF(\bfk)|^2$
\begin{equation}
  P(\bfk)=|\calF(\bfk)|^2-P_{shot}.
\end{equation}
Using the two-point functions of $n(\bfs)$ and $n_s(\bfs)$ 
(see equation~(2.1.5) in FKP), we can compute the expectation
value of $\langle|\calF(\bfk)|^2\rangle$, which yields
\begin{equation}
  P_{shot}=(1+\alpha){\int dz (dN/dz)\over \int dz W(z)}
\end{equation}
and 
\begin{equation}
  \langle P(\bfk) \rangle= {\int dz W(z) P^a(\bfk,z)\over \int dz W(z)},
\label{PLC}
\end{equation}
where we defined $W(z)=(dN/dz)^2(dz/ds)/s^2$ by using the observable
quantity $dN/dz$, the number of sources per unit redshift and unit solid angle,
instead of $\barn(s)$, which are related by $dN=dss^2\barn(s)$,
$P^a(\bfk,z)$ is the local power spectrum defined on a constant time 
surface of the redshift $z$. 
In deriving (\ref{PLC}), we adopted the distant observer approximation,
$|\bfs_1-\bfs_2|<<|\bfs_1+\bfs_2|$. Under this approximation, the correlation 
function is related with $P^a(\bfk,z)$ as
\begin{eqnarray}
\xi(\bfs_1,\bfs_2)=(2\pi)^{-3}\int d{\bfk} 
P^a\bigl(\bfk,z(|\bfs_1+\bfs_2|/2)\bigr) e^{i\bfk\cdot(\bfs_1-\bfs_2)}.
\end{eqnarray}
In a similar way, the variance of the power spectrum 
(covariance matrix) can be approximately estimated
\begin{eqnarray}
  \langle \Delta P(\bfk)\Delta P(\bfk')\rangle=
  {2 (2\pi)^3\over \kappa}\delta(\bfk-\bfk'),
\label{DP}
\end{eqnarray}
where $\Delta P(\bfk)=P(\bfk)-\langle P(\bfk) \rangle$, and
\begin{eqnarray}
\kappa=\Delta\Omega \int ds s^2 \barn(s)^2 =\Delta\Omega \int dz W(z)
\end{eqnarray}
with the solid angle of a survey area $\Delta \Omega$.
Equation (\ref{PLC}) is the power spectrum including 
the light-cone effect, which is incorporated by 
averaging the local power spectrum $P^a(\bfk,z)$ 
over the redshift with the weight factor $W(z)$. 
The usefulness of the power spectrum (\ref{PLC}) is 
discussed by comparing with results of the 2QZ group (Paper I).

In general, the Fisher matrix element is defined by 
\begin{eqnarray}
F_{ij}=\biggl\langle{- \partial^2 \ln L\over \partial
\theta_i\partial\theta_j}\biggr\rangle,
\end{eqnarray}
where $\theta_i$ denotes the parameters and $L$ is the 
probability distribution function of a data, given the
model parameters. For simplicity, we adopt the approximation 
of the Gaussian probability distribution function for $\Delta P(\bfk)$
\begin{eqnarray}
  L\propto\exp\Bigl[ -{1\over2}
  \int  d\bfk \int d\bfk'\Delta P(\bfk) C(\bfk,\bfk')^{-1}
  \Delta P(\bfk')\Bigr],
\label{defL}
\end{eqnarray}
where the matrix $C(\bfk,\bfk')^{-1}$ is the inverse 
of the covariance matrix 
$\langle \Delta P(\bfk)\Delta P(\bfk')\rangle$.
Following the work by Tegmark~(1997), 
the Fisher matrix element reduces to
\begin{eqnarray}
  &&F_{ij}={\kappa\over 4\pi^2} \sum_{l=0,2,\cdots} (2l+1)
  \int_{k_{\rm min}}^{k_{\rm max}}
  dk k^2 {\partial \langle P_l(k) \rangle\over \partial \theta_i}
  {\partial \langle P_l(k) \rangle\over \partial \theta_j},
\nonumber
\\
&&{}
\label{Fij}
\end{eqnarray}
where 
\begin{eqnarray}
  \langle P_l(k) \rangle=\int_0^1 d\mu {\cal L}_l(\mu) 
  \langle P(\bfk) \rangle,
\label{Pl}
\end{eqnarray}
and ${\cal L}_l(\mu)$ is the Legendre polynomial of the $l$-th order,
and $\mu$ denotes the directional cosine between the line-of-sight 
direction and the wave number vector. Note that 
$\langle P(\bfk) \rangle$ is a function of $k$ 
and $\mu$ under the distant observer approximation.
This allows $P^a(\bfk,z)$ in equation (\ref{PLC}) 
to be written $P^a(k,\mu,z)$, which we model as 
(Ballinger et~al. 1996)
\begin{eqnarray}
  &&P^a(k,\mu,s)={1\over \cpara\cperp^2}
\nonumber\\
  &&{\hspace{0.5cm}}\times
  P_{QSO}\Bigl(\qpara\rightarrow{k\mu\over\cpara},
  \qperp\rightarrow{k\sqrt{1-\mu^2}\over\cperp},z\Bigr),
\label{PaPQSO}
\end{eqnarray}
where $P_{QSO}(\qpara,\qperp,z)$ is the QSO power spectrum,  
$\qpara$ and $\qperp$ are the wave number components parallel 
and perpendicular to the line-of-sight direction associated
with the comoving coordinate of the real universe, 
$\cpara$ and $\cperp$ are defined by
$\cpara=dr(z)/ds(z)$ and $\cperp=f_K(r(z))/s(z)$, respectively, 
with the comoving distance $r(z)$ and the angular 
diameter distance $f_K(r(z))$ of the real universe. 
The explicit expression of $r(z)$ is presented by 
equation (\ref{defr}).
We model the QSO power spectrum of the spatial 
distribution incorporating the linear distortion 
(Kaiser 1987, see also Yamamoto Nishioka \& Suto 1999), 
as follows:
\begin{eqnarray}
  P_{QSO}(\qpara,\qperp,z)=
  \biggl(1+{f(z)\over b(z)}{\qpara^2\over q^2}\biggr)^2
  b(z)^2 P_{\rm mass}(q,z),
\label{PQSO}
\end{eqnarray}
where $f(z)=d\ln D(z)/d\ln a(z)$ with the linear growth rate 
$D(z)$ and the scale factor $a(z)$, $q^2=\qpara^2+\qperp^2$, 
$b(z)$ is the scale independent bias factor, 
and $P_{\rm mass}(q,z)$ is the CDM mass power spectrum.
The term in proportion to $f(z)$ describes the linear
distortion due to the peculiar velocity on large scales.  
We work within the linear theory of density perturbations because
we consider the large scale QSO clustering statistics. 
In the present paper, we assume the Harrison-Zeldovich
initial mass power spectrum and adopt the fitting 
formula of the cold dark matter transfer function by
Eisenstein and Hu (1998), which is robust
even when the baryon fraction is large.

Equation~(\ref{Fij}) is an extension of the previous result 
by Tegmark (1997). In the limit that the linear 
distortion, the geometric distortion, and the light-cone
effect are switched off, equation~(\ref{Fij}) reproduces the 
result by Tegmark (1997). 
In the investigation in the present paper, the light-cone effect 
and the geometric distortion effect are particularly important. 
For a sample of a narrow range of the redshift, the light-cone 
effect can be negligible, however, it should be properly taken
into account in the QSO spatial power spectrum analysis of 
the 2QZ sample and the SDSS sample, which consist of
sources in the wide range of the redshift. 
In general, the geometric distortion becomes substantial for 
high-redshift samples. On the other hand, the linear distortion 
effect is rather minor in our investigation. It is well recognized 
that the linear distortion effect is substantial and problematic 
in the Alcock \& Paczynski's geometric test (e.g., Ballinger et~al. 
1996). It is the fact that the anisotropic component of the power spectrum 
is sensitive to the linear distortion effect even at the high 
redshift. However, as we show in the next section, the dominant 
contribution constraining the cosmological parameters 
comes from the isotropic part of the power spectrum.
The linear distortion effect only increases the overall
amplitude of the isotropic part of the power spectrum in 
the scale independent bias model. 
This is the reason why the linear distortion is the minor 
effect in our investigation.
This situation is in contrast to case of low redshift galaxy 
samples, in which the light-cone effect and the geometric 
distortion are negligible, while the linear distortion is 
rather important.

\section{Results}

In general, the comoving distance is written, 
\begin{eqnarray}
 r(z)=\int_0^z {dz'\over H_0
[\Omega_m(1+z')^3+\Omega_K(1+z')^2+\Omega_x(1+z')^{3(1+w_Q)}]^{1/2} },
\label{defr}
\end{eqnarray}
where the Hubble parameter is $H_0=100~h{\rm km/s/Mpc}$,
$\Omega_m$, $\Omega_K$, and $\Omega_x$ are the cosmological 
parameters of the densities and curvature, two of which are 
independent by virtue of the relation $\Omega_m+\Omega_K+\Omega_x=1$, 
and we introduced the constant cosmic equation of state parameter, $w_Q$.
In the present work, we focus on the four independent parameters 
$\Omega_m$, $\Omega_b$, $\Omega_K$ and $w_Q$. Here 
$\Omega_m$ is the sum of the CDM and the baryon components.
Following an analysis by the 2QZ group~(Hoyle et~al. 2002a),
we choose $s(z)$ to be the comoving distance with 
$\Omega_m=0.3$, $\Omega_K=0$, and $w_Q=-1$:
\begin{eqnarray}
 s(z)=\int_0^z {dz'\over H_0
[0.3(1+z')^3+0.7]^{1/2}}.
\label{defs}
\end{eqnarray}
Our results does not significantly depend on this choice of $s(z)$.
For $dN/dz$ in the present paper, we adopt the recent result
reported by Schneider et~al. (2002) with the first SDSS QSO catalog. 
The QSO clustering bias in the SDSS sample has not been 
well investigated at present. However, Croom et~al. reported 
on the time-evolution of the QSO bias by analyzing the 
10k catalog of the 2QZ Survey (2001). According to 
them, the clustering amplitude does not show significant time 
evolution. For modeling the bias, following (Calvao et~al. 2002), 
we first adopt the form
\begin{equation}
  b(z)=1+(b_0-1)D(z)^{-1.7},
\label{bzfirst}
\end{equation}
and determined the constant $b_0(=1.2)$ so that the power 
spectrum best matches the 2QZ result~(Hoyle et~al. 2002a).
For the linear growth rate of the model with $w_Q\neq-1$, 
we adopted the fitting formula developed by Wang and 
Steinhardt (1998). Thus we vary the linear growth rate
depending on cosmology. However, our result is not significantly
altered by this modelling because the linear distortion effect 
is not substantial as described in the previous section.

Figure 1 shows the best statistical errors,
$\Delta \theta_i=1/F_{ii}^{1/2}$, some of which are 
normalized by the target parameters (the parameters 
supposed to be true), $\Omega_m\tr=0.28$, $\Omega_K\tr=0$, 
$\Omega_b\tr=0.04$ and $w_Q\tr=-1$, 
as functions of $k_{\rm max}$ in equation~(\ref{Fij}).  
We fixed $k_{\rm min}=0.005~h {\rm Mpc}^{-1}$ and 
assumed $\pi$ steradian of the survey area.
The solid and dotted curves in each panel are the 
$l=0$ and $l=2$ components of $F_{ij}$, respectively.
The dashed curve corresponds to the case that both 
components are summed up. Contribution of the higher
modes $l\ge 2$ to the Fisher matrix element is small.
Thus the contribution constraining the cosmological 
parameters is dominated by the component $l=0$. 
The component $l=2$ corresponds to the constraint 
from the anisotropic power spectrum $\langle P_2(k)\rangle$, 
while $l=0$ corresponds to the constraint from the isotropic 
power spectrum $\langle P_0(k)\rangle$.
The geometric distortion effect causes anisotropies 
of the power spectrum, and the Alcock-Paczynski's geometric
test utilizes this fact. The geometric distortion effect
also alters the shape of the power spectrum $\langle P_0(k)\rangle$
due to the scaling of the wave number, equation (\ref{PaPQSO}).
Therefore the constraint on the cosmological parameters in the 
present paper mostly relies on the latter effect that the shape 
of the isotropic power spectrum changes due to the geometric distortion.

Figure 2 shows the best statistical errors,
$\Delta \theta_i=1/F_{ii}^{1/2}$, as function of 
$k_{\rm min}$ with fixed $k_{\rm max}=1~h{\rm Mpc}^{-1}$. 
The target parameters and the meaning of the curves are 
same as those of Figure 1. 
These two figures indicate that the Fourier modes 
of $k \simlt 0.01 h{\rm Mpc}^{-1}$ and 
$k \simgt 0.1 h{\rm Mpc}^{-1}$ are not
significantly relevant to the integration 
for the Fisher matrix elements.
Thus the Fourier modes of $0.01h{\rm Mpc}^{-1}\simlt k 
\simlt 0.1 h{\rm Mpc}^{-1}$ dominate over 
the constraint on the cosmological parameters.
This is a contrast to the case of the galaxy 
sample (Tegmark 1997). This difference traces back to 
the shot-noise dominance in the QSO sample.  
This result also suggests the validity of our assumption 
that nonlinear effect of density perturbations is 
small because the nonlinearity is influential 
on scales $k\simgt 0.1 h{\rm Mpc}^{-1}$. 

In the present paper, we have not considered the effect of 
the window function depending on the shape of a survey volume. 
The effect of the window function on the power spectrum
for the 2QZ sample was investigated in the literature 
(Hoyle et~al. 2002a). 
According to their result, the 2QZ power spectrum might be 
affected for $k \simlt 0.02 h{\rm Mpc}^{-1}$. 
Though a definite investigation using a mock sample is 
needed, for the case of the SDSS QSO sample, 
it might be expected that the effect of the window
function appears on the larger length scales (smaller $k$) than 
that of the 2QZ power spectrum. From figures 1 and 2, 
as long as the window function effect in the SDSS QSO sample 
appears on the scale $k \simlt 0.01 h{\rm Mpc}^{-1}$, 
the estimation of the errors is not altered significantly.

We note that our result might depend on the modelling 
of the bias. This is because the Fisher matrix element 
is in proportion to the amplitude of the power spectrum 
of the QSO distribution.
Figure 3 shows the same as Figure 1, but with adopting the bias model 
$b(z)=b_0D(z)^{-1}$, in which the amplitude of the QSO 
clustering is almost constant as a function of  the redshift 
$z$. Here we set $b_0=0.8$ so as to best match the 
2QZ power spectrum (see also Paper I).
The best statistical errors are worse than those of Figure~1, 
though the difference is not large.
The reason of the smallness of the difference is explained 
as follows:
We normalized the power spectrum so as to match the 
2QZ power spectrum, in which the 2dF QSO sample is 
distributed in the range of redshift $0.3\leq z \leq 2.3$.
The difference is the contribution from the higher redshift
$z\simgt2$.  
The bias model (\ref{bzfirst}) has larger amplitude at 
the higher redshift, which yields larger amplitudes
of the power spectrum and the Fisher matrix elements.


In the above investigations, we considered the simple bias 
models, whose amplitude we determined so that the power spectrum
best matches the 2QZ result. This method might estimate the 
errors smaller. Then we also investigated the best 
statistical errors marginalized over parameters of the bias. We here 
consider the bias model parameterized by $b_0$ and $\nu$, as follows:
\begin{equation}
  b(z)=1+{(b_0-1) \over D(z)^{\nu}},
\label{bzsecond}
\end{equation}
then marginalize the probability function with respect to 
$b_0$ and $\nu$.  
Explicitly, we considered the space of the parameters 
$(\Omega_m,~b_0,~\nu)$ or $(w_Q,~b_0,~\nu)$ , and consider 
the Fisher matrix element marginalized by integrating  
the probability function with respect to $b_0$ and $\nu$. 
This process actually reduces the accuracy with which the 
cosmological parameters can be determined. For example, 
in the case with the target parameters $b_0\tr=1.2$
and $\nu\tr=1.7$, we obtained the relative error
$\Delta \Omega_m/\Omega_m \simeq 0.4$ and 
$\Delta w_Q \simeq 0.7$, where we assumed the same
target parameters for the cosmological models as those 
of Figure 1. Similarly, we have $\Delta \Omega_m/\Omega_m \simeq 0.35$ and 
$\Delta w_Q \simeq 0.5$ for the target parameters $b_0\tr=1.5$
and $\nu\tr=1$. We also have 
$\Delta \Omega_m/\Omega_m \simeq 0.25$ and 
$\Delta w_Q \simeq 0.35$ for $b_0\tr=1.5$ and $\nu\tr=1.7$.

\section{Degeneracy}
In general the feasibility of determining the cosmological
parameters is restricted by degeneracies in the parameter
space. Actually many possible parameters can appear 
in generic cosmological model. Here we demonstrate
some aspects of the degeneracies in the cosmological parameters. 
In Figure 4 we plot two dimensional confidence contours, where 
we assumed the same QSO sample and the bias model as those of 
Figure 1, $\Omega_m\tr=0.28$, 
$\Omega_\Lambda\tr=0.72$, $\Omega_b\tr=0.04$, $w_Q\tr=-1$, 
and the Harrison-Zeldovich spectrum $n=1$. 
Each panel shows the confidence contours on the
two parameters, i.e., $\Omega_m$ and the other 
parameter labeled in each panel, with assuming the 
rest parameters to be fixed. Thus this figure shows 
the degeneracy between $\Omega_m$ and other parameters. 
From this figure it is clear that the uncertainty about $\Omega_b$ 
is problematic when determining $\Omega_m$.
It also shows that the degeneracy between $\Omega_m$ 
and $n$, the index of the 
initial power spectrum, is relatively large. 
Thus the degeneracy between these parameters 
will be problematic in terms of determining all the parameters
simultaneously. 
Nevertheless a useful constraint will be obtained 
from the QSO sample by combining it with constraints from
Big Bang Nucleosynthesis, CMB anisotropies, Ia 
supernova, weak gravitational lensing and other 
observations. Then we can emphasize that the QSO sample 
provide a chance of a worthy complimentary test for the
cosmological models.



\section{Discussions and Conclusions}

In this paper we have assessed the feasibility of measuring 
the cosmological parameters by means of the power spectrum analysis 
of the SDSS QSO sample, with an emphasis on a unique observational
probe of the Universe.  
If tight constraints are provided, the result will have
implications for testing the cosmological principle and 
inhomogeneous cosmological models.
The best statistical errors are estimated using the 
Fisher matrix approach, which is a simple extension 
of the previous works for galaxy samples. 
For example, the best statistical errors are at $10\%$
level for $\Omega_m$ and $\Omega_K$, a few times $10\%$ 
for $w_Q$ and $\Omega_b$. These errors can increase to 
several  $\times 10\%$, depending on the bias 
amplitude, when marginalizing over the parameters of the bias. 
Thus our result shows that the SDSS QSO sample might have 
a useful potential as a test of cosmological model.
But, this does not mean that we have shown the capability 
of the Alcock-Paczynski test using the anisotropies of 
the power spectrum, because the constraint is almost from 
the isotropic component of the power spectrum.
The cosmic degeneracy limits the capability of the 
method in determining all 
parameters with the best statistical accuracy simultaneously. 
However, the point is the uniqueness of 
the method as a probe, and the result would be useful 
for more robust understanding of the Universe. 

Finally we mention comparisons with other works related with 
the present paper.
Ballinger et~al. (1996) used a method similar 
to the present paper, to predict a constraint from the 2dF QSO
clustering statistics. 
Outram et~al. (2001) have applied their method to the 2QZ sample.
The 1-$\sigma$ error of $\Omega_m$ by Outram et~al. (2001) 
is consistent with our result obtained by applying our method 
to the 2QZ sample.
Very recently, Matsubara and Szalay (2002) have also reported 
a similar investigation based on an eigenmode analysis, 
while our method relies on the power spectrum analysis. 
Concerning the QSO sample, as we have discussed, the number 
density in proportion to $dN/dz$ and the modelling of the 
bias $b(z)$ can be sensitive to the final results. 
These modellings and the choice of the cosmological 
redshift space $s(z)$ are also different.

%

\vspace{5mm}
{\it Acknowledgments:}~
This work was done during the stay at 
Max-Planck-Institute for Astrophysics (MPA),
supported by fellowships for Japan Scholar 
and Researcher abroad from Ministry of Education, Science 
and Culture of Japan. 
The author thanks Prof.~S.~D.~M.~White, 
Dr.~H.~J.~Mo, Dr.~P.~A.~Popowski, 
and the people at MPA
for their hospitality and useful comments and conversations.
He is grateful to Prof.~K.~Tomita, Fiona Hoyle and H. Nishioka
for useful conversations and communications  
related to the topics of the present paper at the early 
stage of this work. 
He thanks Prof. Y. Kojima for encouragement.
He also thanks anonymous referee of MN for useful comments,
which helped improve the manuscript.

\newpage

\begin{figure}
\begin{center}
    \leavevmode
    \epsfxsize=15cm
    \epsfbox[20 150 600 720]{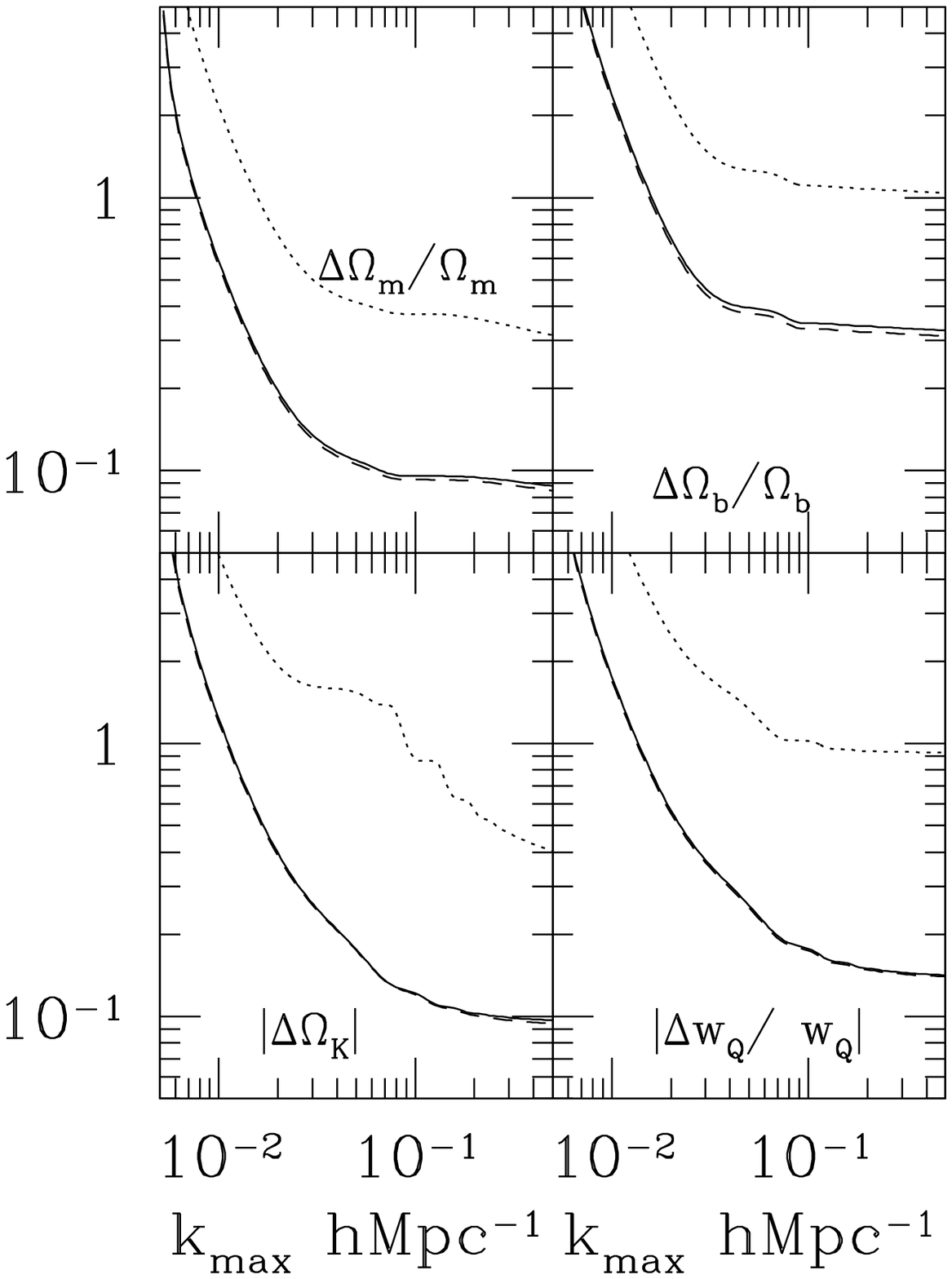}
\end{center}
\caption{The best statistical errors 
$\Delta \theta_i(=1/F_{ii}^{1/2})$ 
as function 
of $k_{\rm max}$, some of which are  normalized 
by the target parameters $\Omega_m\tr=0.28$, 
$\Omega_K\tr=0$, $\Omega_b\tr=0.04$ and $w_Q\tr=-1$.
The solid and the dotted curve in each panel corresponds
to $l=0$ and $l=2$ component, respectively. The dashed
curve is the sum of the both. We adopted the solid angle
of the survey area,  $\Delta \Omega=\pi$~steradian.}
\label{fiburea}
\end{figure}
\begin{figure}
\begin{center}
    \leavevmode
    \epsfxsize=15cm
    \epsfbox[20 150 600 720]{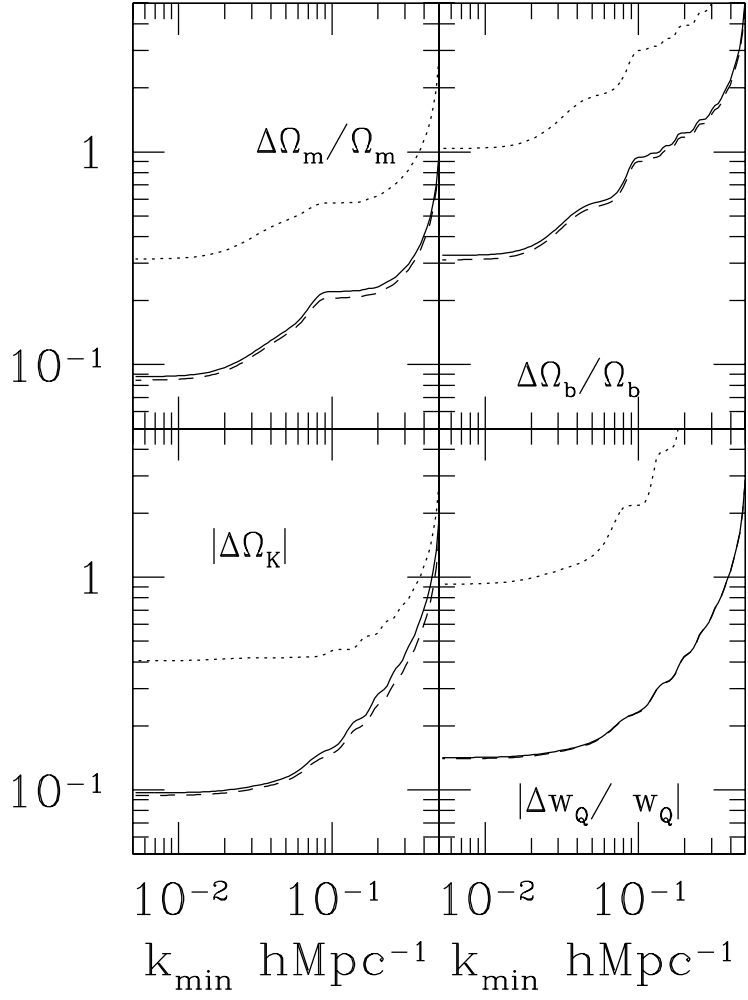}
\end{center}
\caption{The best statistical errors 
$\Delta \theta_i(=1/F_{ii}^{1/2})$ as function of $k_{\rm min}$.
The target parameters and the meanings of the curves are
same as those of Figure 1. Here we fixed 
$k_{\rm max}=1~h{\rm Mpc}^{-1}$.}
\label{figureb}
\end{figure}
\begin{figure}
\begin{center}
    \leavevmode
    \epsfxsize=15cm
    \epsfbox[20 150 600 720]{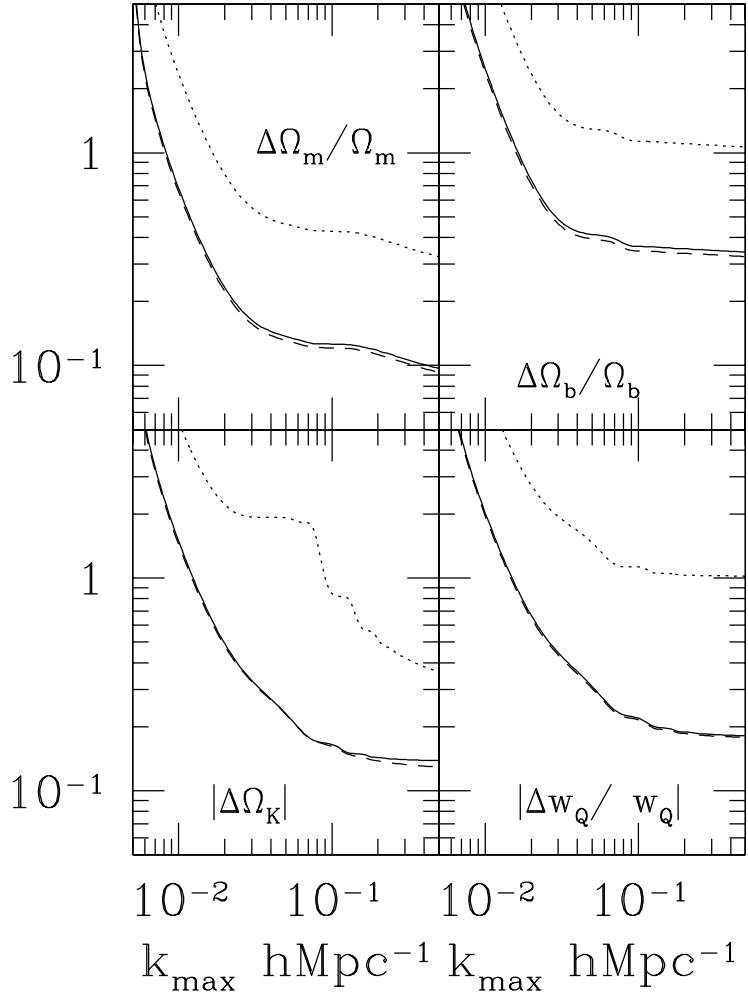}
\end{center}
\caption{Same as Fig.~1 but with the bias model of the form 
$ b(z)=b_0D(z)^{-1}$, where $b_0(=0.8)$ is used.}
\label{figurec}
\end{figure}
\begin{figure}
\begin{center}
    \leavevmode
    \epsfxsize=15cm
    \epsfbox[20 150 600 720]{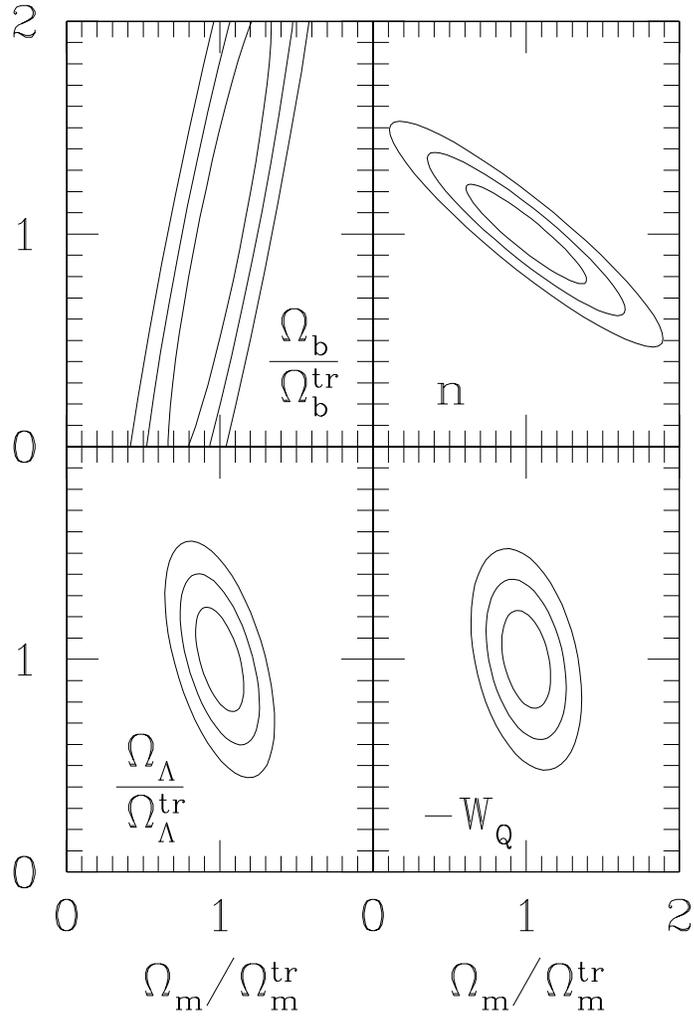}
\end{center}
\caption{Confidence regions.
We have chosen the target parameters as 
$\Omega_m\tr=0.28$, $\Omega_\Lambda\tr=0.72$, 
$\Omega_b\tr=0.04$, $w_Q\tr=-1$ and $n\tr=1$. 
In the panels, we used the superscript $`\rm tr'$ to 
denote the target parameters explicitly.  
Each panel shows the confidence contours on the two 
parameters, $\Omega_m$ and each of the other parameters,
which is labeled in the panel, assuming that the rest 
parameters are fixed.
The curves are the contours of the $68\%$, $95\%$
and $99.7\%$ confidence contours.
Axes are normalized by the target parameter.}
\label{fibured}
\end{figure}

\end{document}